\documentclass[11pt]{article}

\usepackage[utf8]{inputenc}
\usepackage[T1]{fontenc}
\usepackage{lmodern}

\usepackage{amsmath, amssymb, amsthm}

\usepackage{algorithm}
\usepackage{algpseudocode}

\usepackage{graphicx}
\usepackage{caption}
\usepackage{subcaption}

\usepackage{hyperref}

\usepackage[numbers]{natbib}
\usepackage{authblk}

\title{Rapid calibration of atrial electrophysiology models using Gaussian process emulators in the  ensemble Kalman filter}

\author[1]{Mariya Mamajiwala\thanks{Corresponding authors: \texttt{mariya.mamajiwala@nottingham.ac.uk}, \texttt{r.d.wilkinson@nottingham.ac.uk}, \texttt{r.h.clayton@sheffield.ac.uk}}}
\author[2]{Cesare Corrado}
\author[1]{Chris Lanyon}
\author[2]{Steven A. Niederer}
\author[1]{Richard D. Wilkinson}
\author[3]{Richard H. Clayton}
\affil[1]{University of Nottingham, School of Mathematical Sciences, Nottingham, NG7 2RD, United Kingdom}
\affil[2]{Imperial College London, National Heart and Lung Institute, London, SW3 6LY, United Kingdom}
\affil[3]{University of Sheffield, School of Computer Science and Insigneo institute, Sheffield, S10 2TN, United Kingdom}

\begin{document}

\maketitle

\begin{abstract}
Atrial fibrillation (AF) is a common cardiac arrhythmia characterised by disordered electrical activity in the atria. The standard treatment is catheter ablation, which is invasive and irreversible. Recent advances in computational electrophysiology offer the potential for patient-specific models, often referred to as digital twins, that  can be used to guide clinical decisions. To be of practical value, 
we must be able to rapidly calibrate physics-based models using routine clinical measurements.
We pose this calibration task as a static inverse problem, where the goal is to infer tissue-level electrophysiological parameters from the available observations. To make this tractable, we replace the expensive forward model with Gaussian process emulators (GPEs), and propose a novel adaptation of the ensemble Kalman filter (EnKF) for static non-linear inverse problems. The  approach yields parameter samples that can be interpreted as coming from the best Gaussian approximation of the posterior distribution. We compare our results with those obtained using Markov chain Monte Carlo (MCMC) sampling and demonstrate the potential of the approach to enable near-real-time patient-specific calibration, a key step towards predicting outcomes of AF treatment within clinical timescales. The approach is  readily applicable to a wide range of static inverse problems in science and engineering.
\end{abstract}

\section{Introduction}
The heart consists of four chambers: the right and left atria, which receive blood returning to the heart, and the right and left ventricles, which pump blood to the lungs and the rest of the body.
In a healthy heart, electrical impulses originating from the sinoatrial (SA) node in the right atrium initiate contraction, and act to maintain a regular and coordinated rhythm. These impulses ensure that the atria and ventricles contract in a synchronised manner, maintaining efficient blood flow throughout the body. Disruption of this coordination may result in abnormally fast, slow, or irregular heart rhythms collectively referred to as cardiac arrhythmias. Atrial fibrillation (AF) is a common type of arrhythmia that is characterised by irregular and often rapid electrical activity in the atria and is becoming increasingly prevalent\cite{linz2024atrial}. The origin of fibrillatory activity is typically in the left atrium (LA). Radiofrequency (RF) catheter ablation of the left atrial tissue is often the recommended treatment for persistent AF. 

Computational models of cardiac electrophysiology have become valuable tools for understanding the mechanisms that underlie normal and abnormal electrical activity in the heart\cite{clayton2011models}. In order to develop personalised computational models of patient-specific atrial electrophysiology, we must estimate unknown model parameters using  patient data. These models have the potential to predict the success of an RF ablation, and hence can be used to guide clinical procedures\cite{niederer2019computational}. These personalised atrial models require two key elements. The first is anatomy, which can be obtained from magnetic resonance imaging (MRI)\cite{boyle2019computationally}, computed tomography (CT), or electroanatomical mapping systems. The second is a map of material properties that may vary across the cardiac tissue. Traditionally, rule-based methods\cite{bayer2012novel,roney2023constructing,jaffery2024review} have been used to assign spatial parameter fields—such as tissue excitability, conduction velocity, or action potential duration—based on anatomical coordinates, imaging-derived features, or physiological gradients. While these approaches are computationally efficient and interpretable, they provide only point estimates. In our probabilistic framework, these rule-based maps can serve as informative priors, which can then be refined using calibration methods that infer patient-specific parameter distributions from clinical data such as voltage maps, conduction velocities, or pacing measurements\cite{nairn2023differences}. This combination leverages the physiological realism of rule-based assignments while enabling uncertainty-aware, personalized model calibration.

When creating or validating models of the LA from electroanatomical mapping data, there are inherent challenges as the data can be noisy, sparse,
and are typically limited to local activation times (LATs) recorded at a few locations within the LA during programmed pacing. Given these data, our objective is to calibrate an electrophysiology model on a patient-specific anatomy using only the measurements that are obtained in a typical routine clinical procedure, ideally in near real-time so that timely guidance can be given during the procedure. 
Our focus is  developing a fast and efficient method for a probabilistic calibration of the tissue parameters from sparse and noisy observations. This is a critical component in the broader goal of predicting the outcome of radio-frequency (RF) ablation through patient-specific electrophysiological modelling, combined with confidence on these predictions.
The tissue parameters that define a patient-specific model vary across the LA and are thus represented as fields. In previous work we have used Gaussian process manifold interpolation\cite{coveney2020gaussian} and latent Gaussian processes\cite{coveney2022calibrating} to represent this spatial heterogeneity, and Markov chain Monte Carlo (MCMC) for calibration\cite{coveney2022calibrating,coveney2021bayesian}. However, these approaches are computationally expensive and require measurements of effective refractory period (as a surrogate for action potential duration (APD)) that are difficult to obtain within clinically acceptable timescales. Here, we  adapt the ensemble Kalman filter (EnKF)\cite{evensen2003ensemble}, a widely used and robust algorithm for inverse problems and rapid system identification domains\cite{moradkhani2005dual,huang2017calibrating,wu2007application}, to achieve this.
While the EnKF was originally developed for dynamic systems with time-series data, it has also shown promise in static inference problems. 
We propose specific adaptations of the EnKF to enable its use when using  Gaussian process emulators, and demonstrate its effectiveness
on a non-linear exemplar problem and for estimating tissue parameters of the LA. The results are also compared with those from MCMC, obtained at much greater computational cost. Finally, we discuss some advantages and disadvantages of the proposed approach, related work, modifications of the approach, and some directions for future research.
\section{Methods}

\subsection*{Our contributions}

We present a novel adaptation of the ensemble Kalman filter to solve  a static inverse problem in close to real time using  Gaussian process emulators. We give an expression for the ensemble Kalman gain matrix for a non-linear measurement function\cite{tang2014nonlinear} in terms of the Gaussian process emulator (GPE) covariance function, and  apply the method to calibrate tissue parameters in the left atrium (LA), and  we demonstrate that the posterior mean recovers the ground truth parameter value with credible intervals.

\subsection*{Calibrating with an emulator}\label{sec:gpe}

Given noisy measurements, $y\in\mathbb{R}^p$, of LATs and/or APDs at 
certain locations on the left atrial surface, we want to estimate unknown tissue parameters $\theta$. We let $\pi(\theta)$ denote our prior belief about $\theta$ before observing the data, and we assume that
\begin{equation}
y=h(\theta)+\epsilon,
\label{eq:y}
\end{equation} where $h(\theta) \in \mathbb{R}^p$ is the simulated LAT/APD at the same set of spatial locations using tissue properties $\theta$. 
We assume that $ \epsilon \sim \mathcal{N}(0,R)$, i.e., we have zero-mean Gaussian observation error.
Our aim is to estimate the posterior distribution of the parameters given the data, $\pi(\theta | y )$.

However, evaluating $h(\theta)$ is computationally expensive, making many standard computational strategies unfeasible in the clinic. All inference must be done using a limited and precomputed ensemble  of function evaluations $\mathcal{E}=\{ \theta_m, h(\theta_m)\}_{m=1}^M$. Given $\mathcal{E}$, we can train a surrogate  model for $h(\theta)$ that can then be used in any analysis \cite{kennedy2001, gramacy2020}. We use GPEs, and assume that  $h(\cdot) \sim GP(m(\cdot), k(\cdot,\cdot))$, where $m(\cdot)$ and $k(\cdot,\cdot)$ are the prior mean and covariance functions. GPEs are closed under Bayesian conditioning, so that $h(.)| \mathcal{E} \sim GP(\bar{m}(\cdot),\bar{k}(\cdot,\cdot))$ where  $\bar{m}$ and $\bar{k}$ are the posterior mean and covariance functions, which are easy to compute from $\mathcal{E}$\cite{williams2006gaussian}.  Note that $m(\theta) \in \mathbb{R}^p$ and $k(\theta, \theta') \in \mathbb{R}^{p\times p}$. We found no benefit from modelling the covariance between different outputs, and so for computational ease assume independent outputs so that 
$k(\theta, \theta')$ is a diagonal matrix, see Supplementary Section \ref{sec:sampling} for details on the training of GPEs.

Once we have replaced the simulator by an emulator, it is  possible to  sample from the posterior distribution $\pi(\theta | y, \mathcal{E})$ using sampling methods such as MCMC. However, due to the inherent serial nature of MCMC, this may still take too long for clinical purposes, as typically we must simulate long chains requiring of the order $10^5$--$10^6$ iterations. 
Our aim is to estimate the posterior in near real-time during a clinical procedure; therefore, computing a fast and robust estimate of tissue properties is essential. To achieve this, we modify the ensemble Kalman filter (EnKF)  to solve the inverse problem, and show that this provides an approach of sufficient speed for practical clinical use.

\subsection*{Background: Ensemble Kalman filter for dynamic problems}\label{sec:background}

The ensemble Kalman filter (EnKF)\cite{evensen2003ensemble} provides a Monte Carlo approximation to the Bayesian filtering problem, i.e., determine the posterior distribution of the state and parameters at time $t$ given measurements upto time $t$, $\pi(X_t,\theta_t| y_{\{1...t\}})$. For combined state and parameter estimation, we assume the following equations hold:
\begin{align}
{\rm d}X_t &= f(X_t,\theta_t){\rm d}t + \sigma_x {\rm d}B_t \label{eqn:process}\\ 
{\rm d}\theta_t &= \sigma_{\theta} {\rm d}W_t \label{eqn:par}\\
Y_t &= h(X_t, \theta_t) + \sigma_y Z_t \label{eqn:meas}
\end{align}
$X_t$ is the state, $Y_t$ an observation, and $\theta_t$ the unknown model parameter at time $t$. 
Equations (\ref{eqn:process}) and (\ref{eqn:meas}) describe  the process and measurement dynamics respectively, and Equation (\ref{eqn:par}) introduces artificial dynamics for the parameters.   ${\rm d}B_t$ and  ${\rm d}W_t$ are independent Brownian motion increments, $Z_t$ is a sequence of independent standard Gaussian random variables, and  $\sigma_x, \sigma_{\theta}$ and $\sigma_y$ are noise intensity matrices (which may or may not be constant with time). To approximate this evolving posterior, the EnKF employs a Monte Carlo approach in which a finite ensemble (collection) of size $N$, consisting of independent samples of the joint state and parameter vectors, $[X^n_0,\theta^n_0]_{n=1}^N$, is propagated and updated over time. Starting from this ensemble, the EnKF proceeds iteratively through two stages as follows:

Predict: $\begin{bmatrix} \tilde{X}^n_{t+1} \\ \tilde{\theta}^n_{t+1}
\end{bmatrix} =  \begin{bmatrix} X^n_{t} \\ \theta^n_{t}
\end{bmatrix}  +  f(X_t^n,\theta_t^n) \Delta t + \sigma_x \Delta B_t^n \;\; ; \;\; n \in \{1, \ldots, N\}$

Update: $\begin{bmatrix} X^n_{t+1} \\ \theta^n_{t+1}
\end{bmatrix}   = \begin{bmatrix} \tilde{X}^n_{t+1} \\ \tilde{\theta}^n_{t+1}
\end{bmatrix}  +K_{t+1} (Y_{t+1} -  h(\tilde{X}_{t+1}^n, \tilde{\theta}^n_{t+1}) ) \;\; ; \;\;  n \in \{1,\ldots, N\}$

The prediction step is performed using explicit Euler-Maruyama integration of the stochastic process dynamics and at the update stage, the particles are given a correction based on the Kalman gain matrix $K$ (when $h(\cdot)$ is non-linear\cite{tang2014nonlinear}) and the innovation vector $ (Y_{t+1} - h(\tilde{X}_{t+1}^n,  \tilde{\theta}^n_{t+1}) )$. 

\subsection*{Proposed method: EnKF for calibration of a static problem with an emulator}\label{sec:enkf}
Equation \ref{eq:y} defines a static calibration problem, with a single observation, no state variable, and constant unknown parameter. The EnKF is not directly applicable, but can be modified for this situation \cite{iglesias2013ensemble}. 
We modify existing EnKF approaches to the scenario where  the observation operator, $h$, is itself an uncertain surrogate for the deterministic simulator.
 
The absence of state dynamics is straightforward to deal with; it may be ignored and the only process model needed is for the evolution of parameters. Equation (\ref{eqn:par})  is an artificial dynamics introduced to allow to exploration of the parameter space as we assimilate information from the data.
If the difference between the posterior distribution and is large, then assimilating the information in a single step (with the static measurement) is challenging as the particles in the ensemble would initially be in the wrong part of parameter space. Instead, we  create an artificial time history of measurements by perturbing the original (already noisy) measurement at every time step. This allows for a smoother transition of the ensemble from prior to posterior. The perturbation of the data at each step is essential so as not to over-count the information available in $y$. The variance of the added noise is  the original noise variance multiplied by the number of iterations of the algorithm (see Supplementary Section \ref{sec:scalesig} for details).  Similar schemes for perturbing the measurements have been used in other work\cite{kovachki2019ensemble, raveendran2011pseudo,raveendran2012pseudo}.

Finally, in order to account for the GPE uncertainty within the EnKF framework, we propose a modification of the ensemble Kalman gain matrix to reflect that we are using a stochastic emulator for the measurement function. Although there is work on using a GPE in the process Equation\cite{lin2024ensemble} \eqref{eqn:process}, there is no variant of the EnKF  that accounts for emulator uncertainty in the measurement Equation \eqref{eqn:meas}. 
The pseudo-code for our modified EnKF is provided in Algorithm \ref{alg:v3}. 
We show numerically that as accuracy of emulator improves, the posterior obtained via Algorithm \ref{alg:v3} converges to the solution of the EnKF for a known non-linear function, demonstrating consistency. See Supplementary Figure \ref{fig:toy}. Note that when $\sigma_{\theta}=0$, the emulator mean function is replaced with the known measurement function, and the emulator variance is set to 0, i.e. $\bar{m}(\cdot)=h(\cdot)$ and $\bar{k}(\cdot,\cdot)=0$, Algorithm \ref{alg:v3} reduces to the usual ensemble Kalman filter for static problems. In the case $\sigma_{\theta}=0$, it has been established\cite{schillings2017analysis} that Algorithm \ref{alg:v3}, with a finite ensemble size and in the limit when time-step tends to 0, yields posterior mean and variance identical to the true posterior, consistent with the behavior of the standard EnKF. We provide an analysis for the effect of non-zero $\sigma_{\theta}$ in Supplementary Section \ref{sec:proof}.

\begin{algorithm}
    \caption{EnKF for Static Calibration with a GPE measurement operator}\label{alg:v3}
    \begin{algorithmic}[1] 
        \State \textbf{Inputs:} Noise covariance matrix $R$, number of iterations $K$, process covariance $\sigma_\theta$, initial distribution $\mathcal{N}_d(\mu_0, \Sigma_0)$, observation vector $Y \in \mathbb{R}^{p}$.
        \State \textbf{Initialise:} Sample parameters $\theta_{0}^n \sim \mathcal{N}_d(\mu_0, \Sigma_0)$ for $n=1, \ldots, N$.
        \For{$k = 0, \ldots,  K-1$}
            \State $\epsilon_y \sim \mathcal{N}_p(\mathbf{0},K  R)$ \Comment{Perturb observation vector}
            \State $Y_{k+1}=Y+\epsilon_y$
            \For{$n = 1, \ldots, N$}
                \State $\epsilon_\theta \sim \mathcal{N}_d(\mathbf{0},I)$
                \State $\tilde{\theta}^n_{k+1} = \theta^n_k + \sigma_{\theta} \epsilon_\theta$ \Comment{Predict $\theta$}
            \EndFor
            \State $P_{k+1} = \frac{1} {\sqrt{N-1}} \left[  \tilde{\theta}_{k+1} - \langle \tilde{\theta}_{k+1}\rangle  \right] $ 
            \State $H_{k+1} =  \frac{1} {\sqrt{N-1}} \left[  \bar{m}(\tilde{\theta}_{k+1})-  \langle \bar{m}(\tilde{\theta}_{k+1}) \rangle \right ] $  
            \State where $\tilde{\theta}_{k+1}, \bar{m}(\tilde{\theta}_{k+1})$ are matrices with $n^{th}$ column $\tilde{\theta}^n_{k+1}$ and $\bar{m}(\tilde{\theta}^n_{k+1})$
            \State $\langle A \rangle$ denotes the matrix with $N$ identical columns, each being the mean of the columns of $A$ 
            \For{$n = 1, \ldots, N$}
                \State $K_{k+1}^n = P_{k+1} H_{k+1}^T (H_{k+1} H_{k+1}^T + R + \bar{k}(\tilde{\theta}^n_{k+1},\tilde{\theta}^n_{k+1}) )^{-1} $ \Comment{Compute Kalman gain}
                \State $\theta^n_{k+1} = \tilde{\theta}^n_{k+1} + K_{k+1}^n [Y_{k+1} - \bar{m}(\tilde{\theta}^n_{k+1})] $ \Comment{Update $\theta$}
            \EndFor
        \EndFor
    \end{algorithmic}
\end{algorithm}

\subsection*{Electrophysiology model}\label{sec:ep}

We model the LA electrophysiology using the modified Mitchell-Schaeffer model \cite{CORRADO201646}. We assume homogeneous isotropic tissue, characterised by a monodomain model given by the following set of coupled partial differential equations


\begin{align}\label{eqn:vm}
    \frac{\partial v_m}{\partial t} &= \nabla\cdot  \left(D \nabla v_m\right) + h \frac{v_m\left(v_m-v_{gate} \right) \left( 1-v_m\right) }{\tau_{in}} - (1-h)\frac{v_m}{\tau_{out}} + F_{stim}\\
    \frac{\partial h}{\partial t} &= \begin{cases}
(1-h)/\tau_{open} & \text{if } v_m \leq v_{gate} \\
-h/\tau_{close} & \text{otherwise }
\end{cases}\label{eqn:h}
\end{align}
where $v_m \equiv v_m(x,t)$ denotes the normalised trans-membrane voltage field over the atrial surface (i.e. $v_m \in [0,1]$),  $h \equiv h(x,t)$ is a gating variable that controls recovery, and   $F_{stim} \equiv F_{stim}(x,t)$ is an externally applied forcing/stimulus. We wish to infer $\theta=(\tau_{in},\tau_{out},\tau_{open},\tau_{close}, D)$, where $\tau_{in},\tau_{out},\tau_{open},\tau_{close}$  are  cell parameters that characterise the four stages of the trans-membrane potential,  and  $D$  is the tissue conductivity. Prior  parameter ranges are summarised in Supplementary Table \ref{tab:tissue}.
Boundary conditions are no-flux across the tissue boundaries to reflect electrical isolation of the tissue, and we use initial conditions that set all cells to resting membrane potential,
and  we fix the  cell excitation threshold at $v_{gate}=0.1$. We use openCARP\cite{openCARP-paper} to solve Equations (\ref{eqn:vm}) and (\ref{eqn:h}) using a finite element method, with a computational mesh  generated from imaging data \cite{doi:10.1161/CIRCEP.121.010253}.  We chose a relatively large LA geometry, since such geometries are more likely to sustain AF, which is our focus, and in previous work\cite{MamajiwalaCinc2024} we showed that parameter sensitivities of tissue outputs are largely independent of geometry. We simulated a paced activation consisting of a stimulus applied in the rectangular region shown in Supplementary Figure \ref{fig:stim}, located near the coronary sinus (CS). We use an S1S2 pacing protocol consisting of 3 S1 stimuli 800 ms apart, followed by a premature S2 stimulus at a 500 ms coupling interval.

We focus on local activation times (LATs), which correspond to the time of rapid depolarisation of the cardiac tissue,
as they are one of the few electrophysiological markers that can be directly recorded during electroanatomical mapping. 
We define the LAT at atrial location $x$, to be the  time at which the voltage at $x$, $v_m(x,\cdot)$, crosses a threshold of $0.75$ from below (i.e., crosses with a positive time derivative). 
In addition to LAT, we also compute the  action potential duration (APD) at each atrial location. We define APD  to be the difference between Local Recovery Time (LRT) and LAT, where LRT is the time at which $v_m$ drops back to 10\% of the maximum amplitude. While APD is not routinely measured in the clinic, it can be approximated from effective refractory period (ERP) measurements, i.e.,  the minimum distance between two subsequent stimuli that fails to propagate. Measuring ERP is complex as it requires repeating the S1S2 protocol for several progressively decreasing values of the coupling interval S2, until failure to propagate. 

Our rationale for including APD in the present study was to examine the potential benefit to calibration afforded by the use of ERP measurements in addition to LAT, so that these benefits can be weighed against the additional cost of making them.

\section{Results}

We assess the efficacy of the proposed EnKF algorithm to accurately and quickly calibrate in non-linear static settings in a toy problem (Supplementary Section \ref{sec:toy}) and in the electrophysiology model, as discussed in the next subsection. The speed of the EnKF approach  allows us to investigate the practical identifiability of the tissue parameters  using different combinations of S1S2 measurements, enabling us to assess their relative value to guide clinical decisions. In the second part of our results, we conducted a pilot study to predict atrial fibrillation post-ablation using calibrated posterior parameters (using S1S2 observations), which represents the motivating clinical application.

\subsection*{Calibration of tissue parameters of the LA}\label{sec:calib:la}
We evaluate the proposed approach for rapid calibration of tissue parameters of the LA based on noisy measurements of LAT and APD. As a first step, we trained Gaussian process emulators (GPEs) of LATs and APDs as a function of the tissue parameters. Prior ranges for the parameters are given in Supplementary Table \ref{tab:tissue}\cite{corrado2018work}. 
Best practice \cite{gramacy2020} suggests the design $\{\theta_i\}_{m=1}^M$ used to generate training data $\mathcal{E}=\{ \theta_m, h(\theta_m)\}_{m=1}^M$ should be space-filling, such as a maximin Latin hypercube \cite{gramacy2020}. 
However, the prior support contains parameter values that produce activation patterns that are not physiologically plausible (for example, APDs larger than 500~ms). Therefore, we designed a series of rejection steps to arrive at the prior support to train the GPE. We first trained a classifier to predict whether a cell model evaluated at $\theta$ results physiologically feasible outputs. Because the cell model (rather than the tissue model) is cheap, we can afford to do sufficient simulation to train a classifier with high predictive accuracy. We then created a space filling design in the prior support, and rejected parameters if the classifier predicts nonphysiological cell predictions. We then perform a second layer of rejection based on tissue simulations performed with openCARP on ARCHER2\cite{beckett2024archer2} with the S1S2 protocol described in the electrophysiology section (some parameter values gave plausible cell simulations, but implausible simulations across the LA).
This resulted in an ensemble of $M=202$ simulations. See Supplementary Section \ref{sec:sampling} for full details.

From the simulations, we extracted the following information:  (i) local activation time (LATs) of the third S1 beat; (ii) LAT of the S2 beat; and (iii) action potential duration (APD) of the S2 beat, which we refer to as S1, S2, and APD henceforth. In a clinical setting, activation maps can make recordings at 100s of points on the left atrium when using a fixed pacing, but  S1S2 pacing protocols tend to only record LAT at a smaller number of locations. To reflect this we used measurements from  15 spatially well-distributed locations on the left atrium (Supplementary Figure \ref{fig:stim}) , based on the universal atrial coordinate (UAC) system\cite{roney2019universal}. The output of the simulator, $h(\theta)$, can thus be considered to be a 45 dimensional vector (3 output types at 15 locations).   We split the ensemble $\mathcal{E}$ into training and test sets, and  trained independent Gaussian process emulators (GPEs) for each output variable. 
The resulting GPEs were validated by assessing their predictive accuracy on the unseen test data. In all 45 cases, the coefficient of determination, $R^2$, which computes the proportion of variance explained, was greater than 95\%. See Supplementary Section \ref{sec:sampling} for full details.

To assess the practical identifiability \cite{wieland2021structural} of the parameters, and how that changes with the different  observation types, as well as  the efficacy of Algorithm 1, we conducted a series of 50 calibration experiments. In each, we selected a ground truth parameter and simulation from the ensemble $\mathcal{E}$,  and added zero-mean Gaussian noise to the outputs to create synthetic measurements. 
Calibration was performed separately for each of the following measurement combinations: (i) S1 only, (ii) S1 and S2, and (iii) S1, S2 and APD. For each instance, the ensemble mean of the proposed method at the final iteration was taken as the calibrated parameter estimate. The results are presented in Figure \ref{fig:par:scatter}. It is clear that $\tau_{in}$ and conductivity, $D$, can be identified from the data - albeit with varying accuracy - irrespective of the type of measurements used for calibration. Other parameters, however, were difficult to estimate with S1 measurements alone. They only become identifiable with the addition of more measurements. This is consistent with global sensitivity analyses of the model\cite{MamajiwalaCinc2024}, where it was shown that $\tau_{in}$ and conductivity are the most influential on the S1 output, and other parameters become important for other outputs such as APD.

To assess the predictive accuracy of the calibrated parameters, we performed S1S2 simulations using the calibrated parameters corresponding to the 150 instances. From these simulations, we extracted the S1, S2 and APD outputs across the entire left atrial (LA) mesh. We then calculated the root-mean-square errors (RMSEs) for the three outputs over the LA mesh (\textasciitilde 300,000) nodes. Figure \ref{fig:err:boxplots} presents the distribution of these RMSEs. We now discuss the three plots in Figure \ref{fig:err:boxplots}. The first plot presents the RMSE in the predicted S1 outputs. This error is highest when calibration is performed using only S1 measurements. Incorporating S2 measurements leads to a modest improvement, while the inclusion of APD measurements does not yield substantial additional gains. 
While the addition of S2 measurements enhances the calibration of other parameters (see Figure \ref{fig:par:scatter}), it also improves the estimation of $\tau_{in}$ and the conductivity, both of which are influential in determining the S1 output. On the other hand, APD measurements primarily enhance the calibration of parameters that have limited impact on S1 output, explaining the minimal reduction in RMSE. The second figure shows the RMSE in the S2 output. Here, the RMSE is notably high when calibration is based solely on S1 measurements. This is expected, as S2 outputs are influenced by the S1S2 coupling interval, which is not captured by the S1 output alone. Once S2 measurements are included in the calibration process, the RMSEs decrease significantly as shown in the middle plot of this figure. As with the S1 case, the addition of APD measurements further refines the calibration of parameters that do not substantially affect S2 outputs, resulting in only a marginal improvement in RMSE. Finally, the third plot displays the RMSE in the APD output. In this case, a clear and consistent reduction in RMSE is observed as more measurement types are incorporated. The APD is governed primarily by $\tau_{in}$, $\tau_{out}$, and $\tau_{close}$, which are not fully captured when using only S1 data. The inclusion of S2 outputs enhances the calibration of these parameters, and the direct use of APD measurements further improves their estimation. This progressive refinement is reflected in the observed reduction in RMSE.

For clarity, we show the LAT maps corresponding to one of the 50 instances using S1+S2+APD measurements in Figure \ref{fig:lat:maps}. It may be observed that the S2 output (4th column) in the first row has the largest error - this is not surprising as this row corresponds to calibration against only S1. All other errors are negligible. For 2 of the 50 ground truths, we show comparison of results from the proposed approach with those via MCMC, see Figure \ref{fig:mcmc:main}. Results for other ground truth instances compare similarly. Metropolis-Hastings is implemented using the emcee package in Python with GPEs. The prior distribution used in MCMC is the same as the initial distribution for the propsed approach. We run 10 independent chains, each with 40,000 samples; removing 10,000 samples as burn-in for each chain and then and applying thinning with a factor of 10, we finally have 30,000 samples for each ground truth. Pair-wise scatter plots are shown in Figure \ref{fig:mcmc:main}. Compared to MCMC or other sequential Monte Carlo methods, the quality of performance of the proposed approach scales more effectively with increasing measurement data. Contrary to expectation, MCMC methods often struggle as more data becomes available. This trend is clearly observed in Figure \ref{fig:mcmc:main}, where the performance of MCMC deteriorates while that of the proposed method improves as more measurements are used. This can be understood intuitively: as the quantity of data increases, the posterior distribution becomes more concentrated, making it harder for proposal samples to fall within high-probability regions. Consequently, a greater proportion of proposals are rejected, leading to poor mixing and requiring significantly longer chains to adequately explore the posterior. In contrast, the proposed approach applies additive corrections to all ensemble members based on the available measurements, enabling efficient assimilation of information and improved convergence. 

The computational cost of the algorithm is $O(NK)$, but the computational time is $O(K)$ if we parallelise. We varied the hyperparameters in Algorithm \ref{alg:v3} to assess robustness to these choices. Specifically, we varied the
ensemble size $N$ from 200 to 500, the measurement noise intensity from 0.01 to 1, and the number of steps $K$ from 20 to 100.  For hyperparameter values in this range, the change in the posterior mean estimate was indistinguishable from the Monte Carlo variability. The computational cost per sample for MCMC also increases with more data because evaluating the likelihood (often via the forward model) is more expensive. On the other hand, adding more measurements does not significantly increase cost per update for the proposed method since the update step involves linear algebra operations (matrix-vector multiplications and inversions), whose cost grows modestly with more measurements. The number of ensemble members usually stays fixed (e.g., 100–500), and all members may be updated in parallel. Furthermore, as more measurements are added, EnKF often converges in fewer iterations because the system is better constrained. Overall, potential for parallelisation and linear update structure make EnKF computationally more scalable with increased data.

\subsection*{Prediction of AF - pilot study}\label{sec:af}

Our objective is to predict whether atrial fibrillation (AF) will be sustained or terminated in a particular patient, for example, after a treatment such as catheter ablation. AF is a complex phenomenon, and it is difficult to collect directly relevant electrophysiology data. Usually, we only have data from S1S2 or similar protocols, and on the basis of this, need to estimate tissue parameters before making a calibrated prediction about AF termination. 
AF initiation in the model can be achieved using various methods, such as cross-field stimulation\cite{pertsov1993spiral} or multiple spiral waves. We used a four-spiral wave initiation protocol based on universal atrial coordinates (UACs) and performed electrophysiology simulations accordingly\cite{roney2019universal}. From the simulation results, we classify each parameter vector, $\theta$, as either leading to sustained AF or to successful termination. 
Averaging with respect to the posterior distribution for the parameters, then gives us an estimate  of whether AF will sustain for a particular patient, given the available data. Explicitly, if $g(\theta)$ is an indicator function taking the value $1$ if AF sustains in the simulation, and is otherwise $0$, then
$$\mathbb{P}(\mbox{AF sustain}\mid y)= \int g(\theta) \pi(\theta \mid y){\rm d} \theta.$$
The challenge is that $h(\theta)$ (the simulator prediction of the S1S2 outcomes) and $g(\theta)$
may have very different sensitivites to $\theta$. Thus, the parameters we can identify from the S1S2 data may not be the parameters we need to know to predict AF termination.

We performed a pilot study to gauge the extent to which this was true, and thus to assess the feasibility of predicting the behaviour of AF from parameters calibrated against S1S2 measurements. 
Our objective was to predict whether AF is terminated within a specified time interval for each posterior sample and compare the predicted outcomes against those from the ground truth parameters. To this end, we chose 2 cases from $\mathcal{E}$, and  for each, we 
selected a representative subset of 200 $\theta$ samples from both the prior and posterior distributions.
AF simulations were performed for all samples, as well as for the corresponding ground truth parameters. The classification outcomes are shown in Figure \ref{fig:mcmc:main}. At least 94\% of the posterior samples correctly predict AF termination across all cases, in alignment with the ground truth outcome, suggesting that despite limitations around identifiability, our proposed approach holds promise for potential clinical applications such as guiding ablation therapy.

\section{Discussion}\label{sec:discussion}

This work builds on and extends our previous approaches \cite{coveney2020gaussian, coveney2021bayesian, coveney2022calibrating}, as well as several established frameworks for Bayesian parameter inference in computationally intensive settings. In particular, our approach relates closely to methods such as Ensemble Kalman Inversion (EKI) \cite{iglesias2013ensemble,iglesias2021adaptive}, the Ensemble Kalman Sampler (EKS)\cite{garbuno2020interacting}, and the broader “Calibrate–Emulate–Sample”\cite{cleary2021calibrate} paradigm. Below, we situate our approach in the context of these methods, highlighting key differences and contributions.

The proposed algorithm can be viewed as a generalisation of Ensemble Kalman Inversion (EKI) \cite{kovachki2019ensemble}, differing in three important aspects. First, it incorporates parameter dynamics via a pseudo-time evolution, whereas standard EKI corresponds to the special case where this evolution noise $\sigma_\theta$ is zero. Second, our method modifies the measurement perturbation strategy by scaling the measurement noise with the number of iterations ($K \times R$), an idea explored in prior work \cite{iglesias2013ensemble,iglesias2021adaptive}. Third, it replaces the deterministic forward model with a GPE, treating the emulator mean $\bar{m}(\cdot)$ as a surrogate for the true measurement operator and including emulator uncertainty through the kernel $\bar{k}(\cdot,\cdot)$. We tested simplified variants of our method that closely resemble EKI by incorporating the first two modifications, but these typically yielded less accurate posterior means and underestimated posterior variances—consistent with known behavior of EKI. Additionally, we experimented with omitting the GPE covariance term in the Kalman gain; while this works well when emulator uncertainty is negligible as compared to the measurement uncertainty, i.e. $k(\cdot,\cdot) \ll R$, it may introduce a bias when the GPE is less accurate.

Our method also diverges from the “Calibrate–Emulate–Sample” framework \cite{cleary2021calibrate}, where emulator training, calibration (typically using the exact forward model), and sampling (often via MCMC) occur sequentially after data acquisition. This workflow, while flexible, is computationally intensive and not well-suited for time-critical applications. In contrast, our approach can be viewed as “Emulate–Calibrate”—we shift the expensive GPE training phase to an offline stage before measurements are available. The inference step then proceeds quickly using GPE evaluations alone, producing approximate posterior samples directly through Algorithm \ref{alg:v3}, without a separate sampling phase. This design makes our method practical for real-time or near-real-time applications.

Finally, our work is conceptually related to the Ensemble Kalman Sampler (EKS) \cite{garbuno2020interacting}, which also aims to approximate posterior distributions using interacting ensemble updates. However, EKS is designed for settings where the forward model $h(\theta)$ is available but its derivatives are intractable. In our case, the forward model itself is expensive to evaluate, motivating the use of a surrogate model. We address this challenge by incorporating a GPE into the inference framework, enabling efficient updates while still targeting a Gaussian approximation of the posterior. Despite the different assumptions about model accessibility, both methods aim to sample from the same target distribution, i.e. a Gaussian approximation of the posterior.

Our proposed approach is particularly well-suited for applications requiring real-time or near-real-time predictions. Most of the computational burden is handled offline, where GPEs are trained over the relevant parameter space. Once trained, the inference step becomes computationally inexpensive, enabling rapid predictions as soon as measurements become available. This makes the method ideal for time-sensitive or clinical scenarios. For the problems considered in this work, a single finite element simulation on ARCHER2 takes approximately 15 minutes. Therefore, calibrating with 1,000 simulations (with an ensemble size of 500 over 20 steps) requires not only approximately 250 hours of HPC compute time, but also significant pre- and post-processing effort. This is because the 500 initial simulations must be completed and post-processed before the next 500 sets of parameters can be generated (with the calibration algorithm) pre-processed and run—introducing substantial overhead costs. In contrast, calibrating with GPEs via Algorithm \ref{alg:v3} takes less than one minute on a standard CPU.

Another strength of the approach is that the ensemble is embarrassingly parallel. Ensemble members can be updated independently, allowing efficient use of computational resources. Convergence is typically achieved within 20–30 iterations, and theoretically, in static settings with a single observation, the algorithm can converge in just one step. This efficient convergence further supports its practical utility in demanding environments. In terms of scalability, the method performs robustly as parameter dimensionality increases. Unlike Sequential Monte Carlo methods, which may suffer from slow mixing and low acceptance rates in high-dimensional spaces—especially in the absence of informative priors—the ensemble Kalman Filter (EnKF) remains computationally feasible. This is due to its additive updates, which do not rely on an accept-reject mechanism. 

Although MCMC provides asymptotically exact samples from the posterior, it often faces practical challenges such as poor mixing or entrapment in regions of high GPE uncertainty (see Figure \ref{fig:mcmc2}). Our numerical experiments show that the proposed approach consistently avoids these pitfalls, converging to the correct mode more reliably. While the EnKF does not yield exact posterior samples but rather samples from a Gaussian approximation, this is often sufficient for parameter estimation tasks—our primary focus in this work.

It is important to emphasise that the accuracy of the calibration process (via any method) is fundamentally tied to the fidelity of the GPE. The proposed algorithm identifies parameters such that the discrepancy between GPE-predicted outputs and the observed measurements aligns with the assumed measurement noise. As illustrated in Supplementary Figure \ref{fig:toy}, even if the inferred parameters exhibit some bias relative to the true values, the measurements predicted by the GPE at those parameters remain unbiased—confirming that the calibration is consistent with the trained emulator. Notably, measurements predicted using the original forward model (not shown) do show bias, attributable to GPE approximation error. This bias diminishes as GPE accuracy improves, reinforcing the need for high-quality surrogate modeling in the offline phase.

\section{Conclusions}
 We have proposed an adaptation of the ensemble Kalman filter to solve a static inverse  problem of the type $y=h(\theta)+\epsilon$. Given noisy measurements, $y$, the objective is to determine the posterior distribution of $\theta$ where $h(\theta)$ is only available as a GPE. With the help of a toy problem, we have shown that Algorithm \ref{alg:v3} converges to the original EnKF solution as the emulator uncertainty goes to zero. We then showed via a sketch proof, that the effect of the pseudo-parameter dynamics on the posterior variance is an inflation by $\sigma_{\theta}\sigma_{\theta}^T t$ relative to the case $\sigma_{\theta}=0$ when the ensemble size goes to infinity and integration time-step goes to zero. The inflation for a finite ensemble size is therefore bound above by $\sigma_{\theta}\sigma_{\theta}^T t$. In practice, however, we have shown through various examples, that this effect is negligible.

\begin{figure}[ht!]
\centering
    \begin{subfigure}{0.4\linewidth}
        \centering
        \includegraphics[height=1.7in]{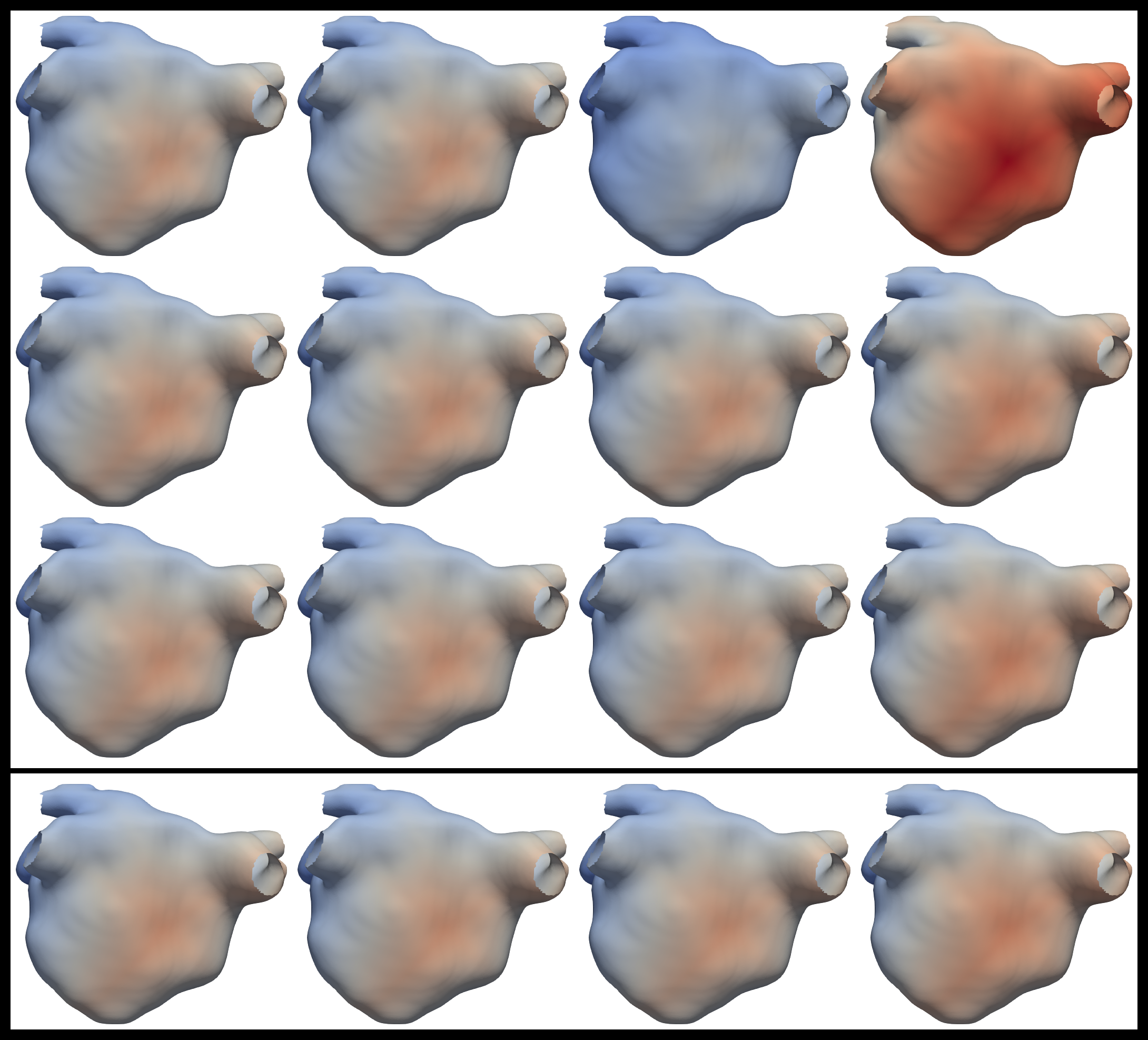}
    \end{subfigure}%
    \begin{subfigure}{0.4\linewidth}
        \centering
        \includegraphics[height=1.7in]{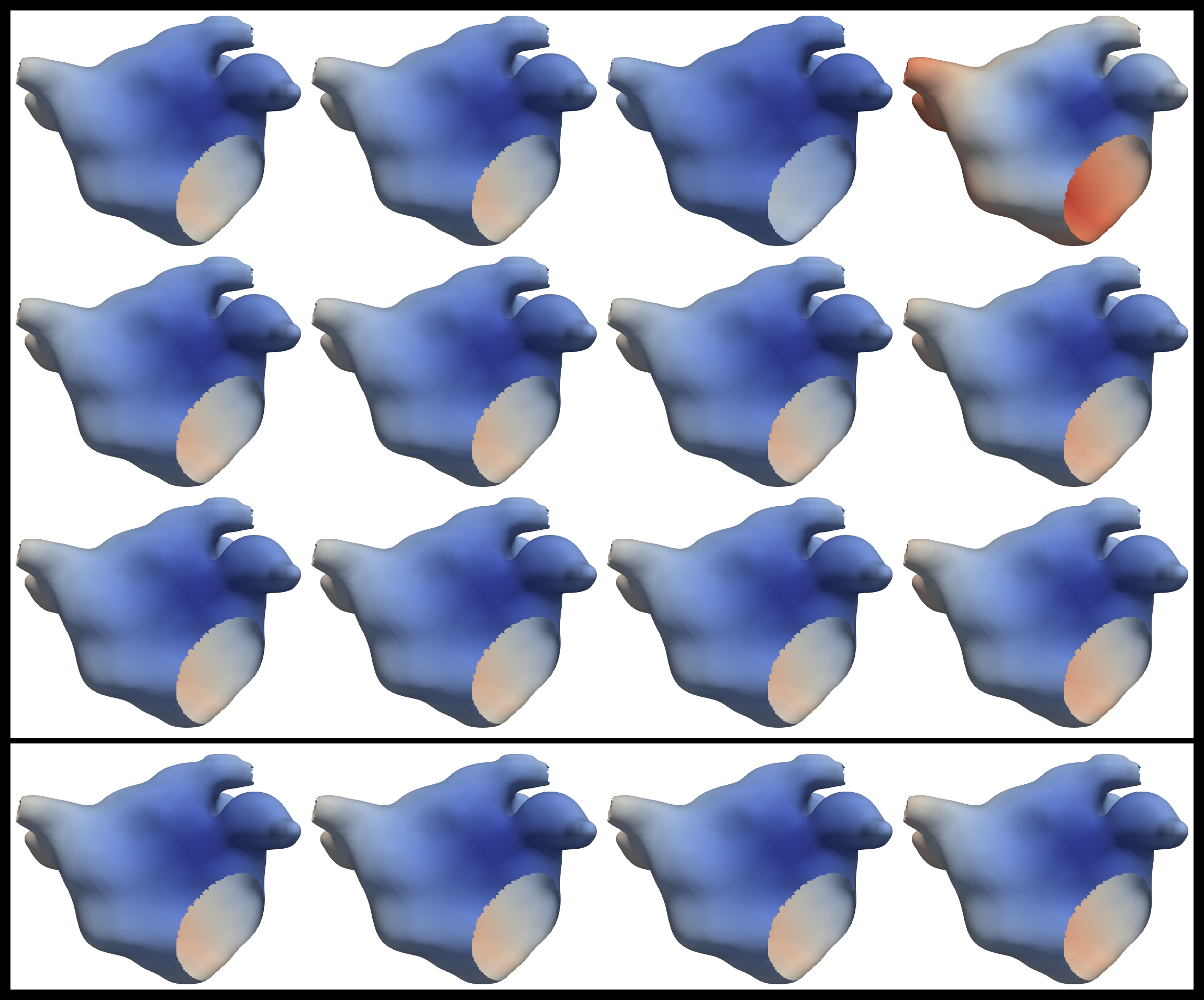}
    \end{subfigure}%
    \begin{minipage}[c]{0.2\linewidth}
        \centering
        \vspace*{-1.7in} 
        \includegraphics[height=1.8in]{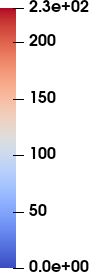}
    \end{minipage}

    \caption{Local activation time maps (in ms). The left and right subfigures show different perspectives. In each subfigure, the four columns correspond to the four LAT maps obtained during the S1S2 simulation. The top three rows show LAT maps simulated with the posterior point estimate of the parameters (i.e., with the ensemble mean after the final iteration of the EnKF) for different combinations of measurements:  S1 only, S1+S2, and S1+S2+APD. The bottom row shows the LAT maps simulated using the  true parameter values.}
    \label{fig:lat:maps}
\end{figure}

\begin{figure}
    \centering
    \includegraphics[width=0.95\linewidth]{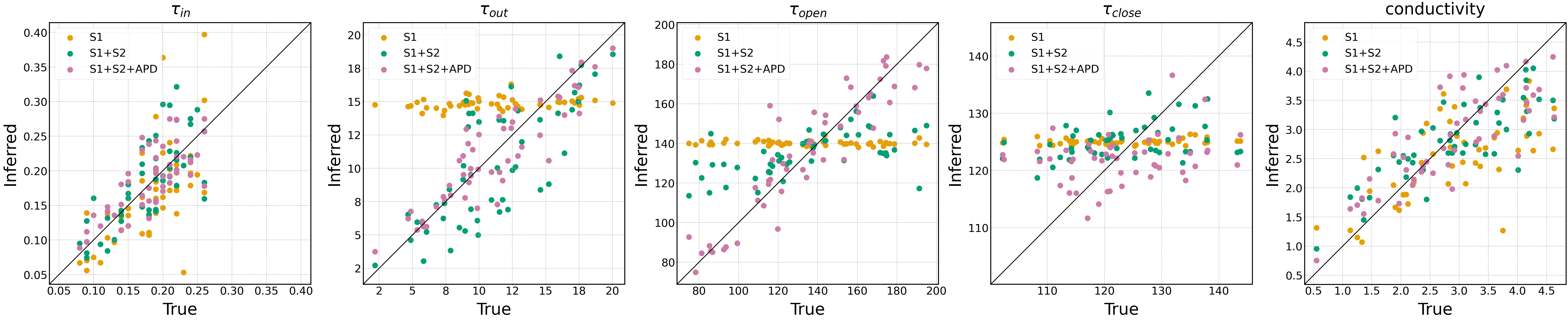}
    \caption{Comparison of ground truth and calibrated parameters for 150 instances of calibration. Each subfigure corresponds to one of the tissue parameters. In each subfigure, the true parameter values are plotted on the x-axis, and the corresponding inferred (calibrated) values on the y-axis. Different measurement types are indicated by distinct colors}
    \label{fig:par:scatter}
\end{figure}

\begin{figure}
    \centering
    \includegraphics[width=0.95\linewidth]{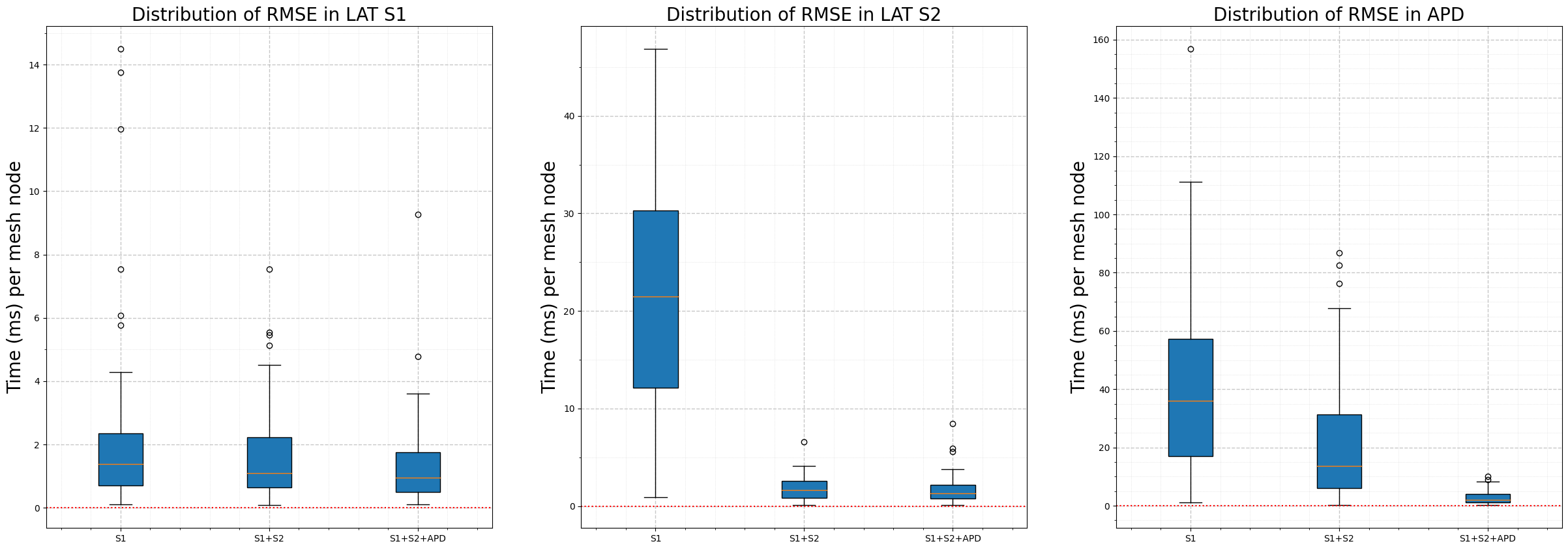}
     \caption{The Figure shows the distribution (over 50 simulations) of RMSEs (over the LA mesh with \textasciitilde 300,000 nodes) of different outputs, viz. RMSEs in LAT S1, LAT S2 and APD in the left, middle and right subfigures respectively. In each subfigure, the three boxplots correspond to 3 calibration scenarios viz., when only S1 (LAT of third S1 beat) measurement is used for calibration, when S2 (LAT of the S2 beat) is used in addition to S1 and when S1+S2+APD (LAT of S1 and S2 beats in addition to the last APD) are used.}\label{fig:err:boxplots}
\end{figure}

\begin{figure}[ht]
\centering
\begin{minipage}{0.75\paperwidth}
    \begin{subfigure}{0.5\textwidth}
    \centering
         \includegraphics[width=\linewidth]{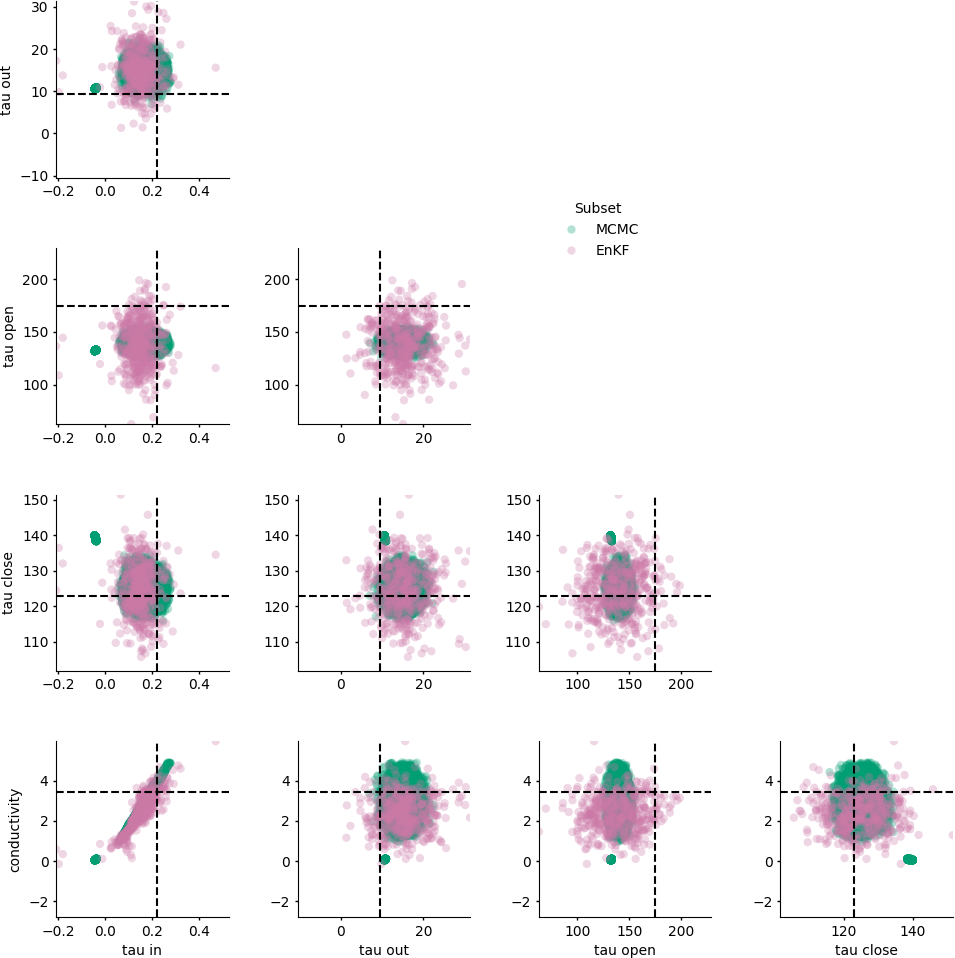}
         \caption{}
    \end{subfigure}
    ~
    \begin{subfigure}{0.5\textwidth}
    \centering
         \includegraphics[width=\linewidth]{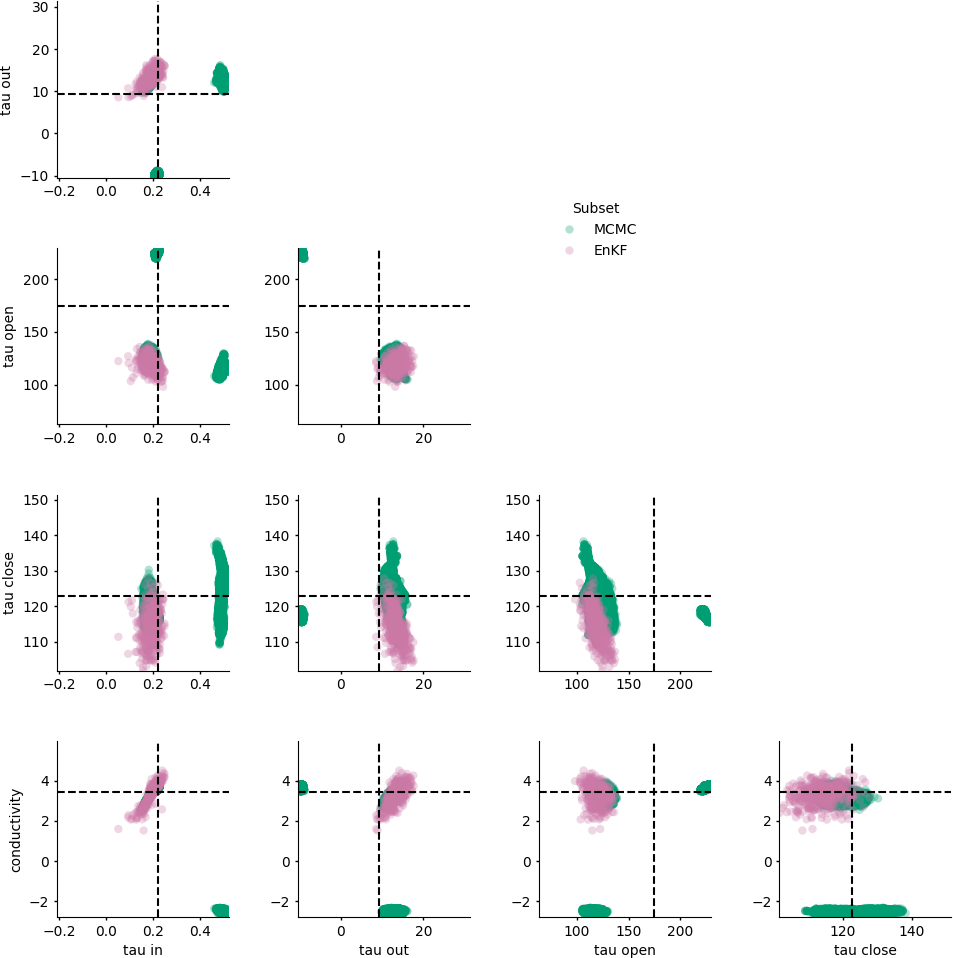}
         \caption{}\label{fig:mcmc2}
    \end{subfigure}
    ~
    \begin{subfigure}{0.5\textwidth}
    \centering
         \includegraphics[width=\linewidth]{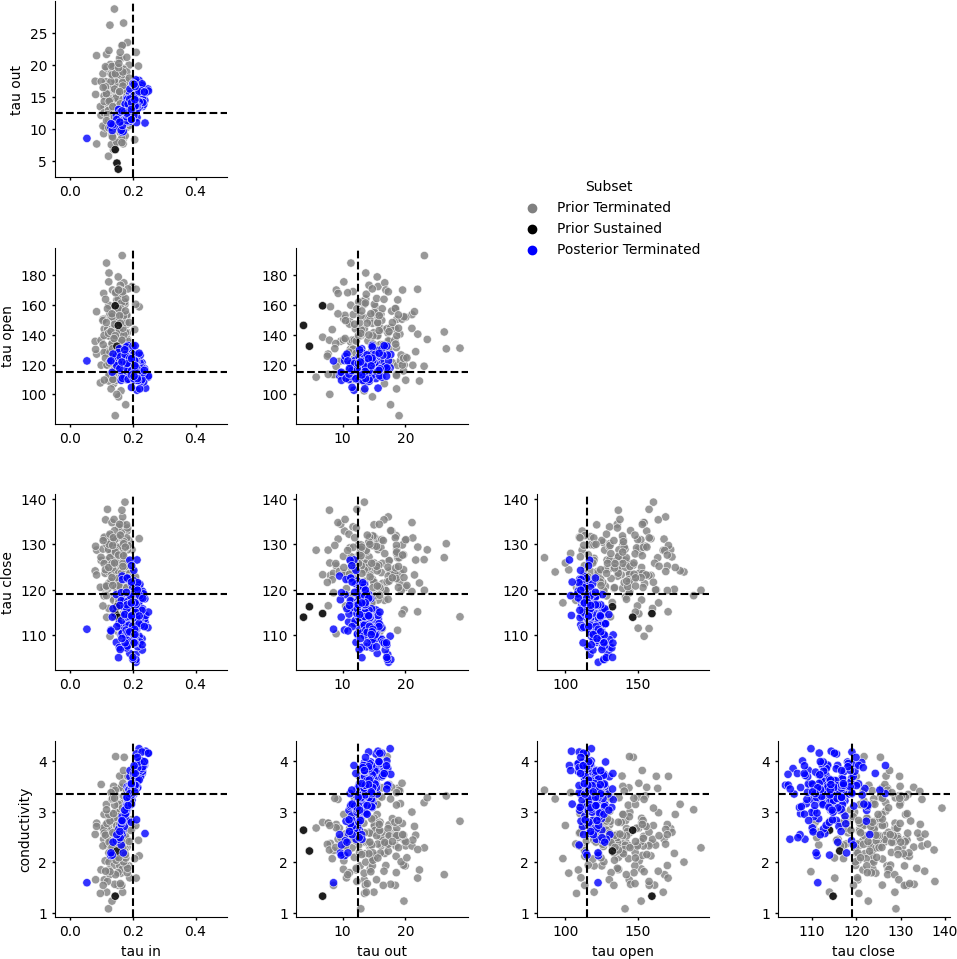}
         \caption{}
    \end{subfigure}
    ~
    \begin{subfigure}{0.5\textwidth}
    \centering
         \includegraphics[width=\linewidth]{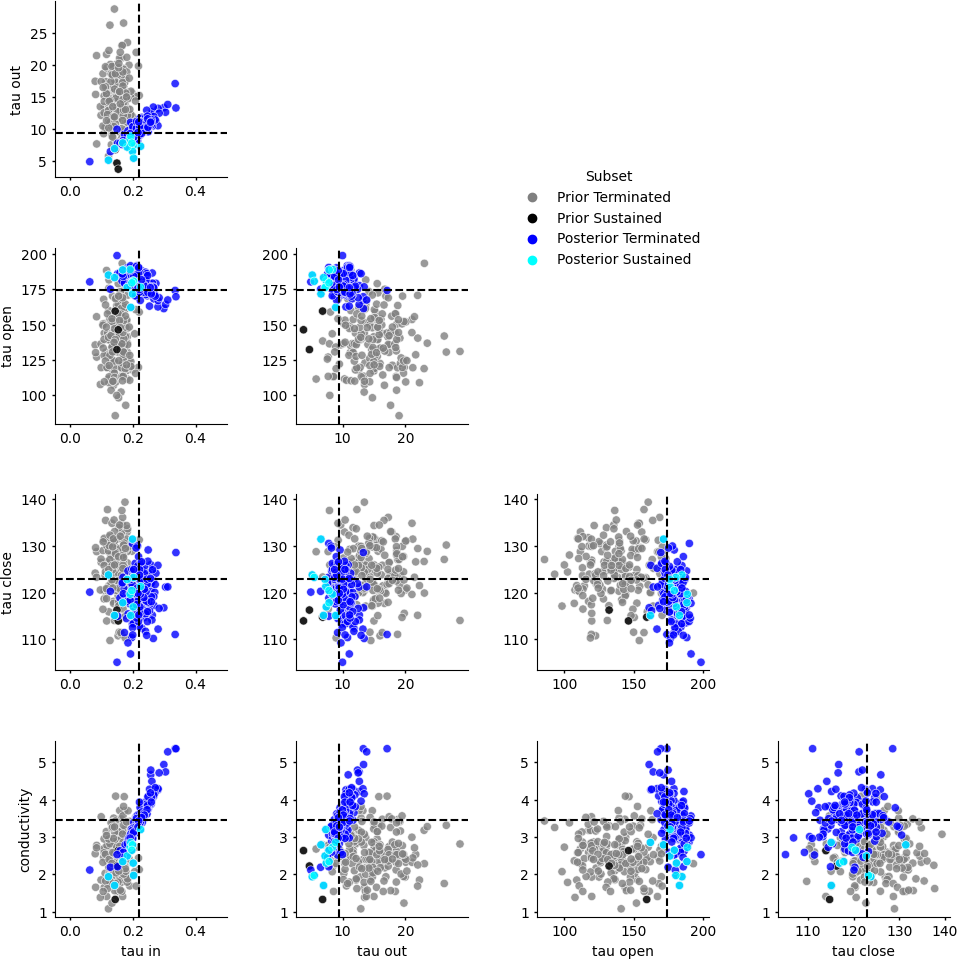}
         \caption{ }
    \end{subfigure}
    
    \caption{Top row: Comparison of posterior distribution of parameters obtained from MCMC and the proposed approach. Subfigure (a) shows results of calibration against only S1 measurements, while (b) shows those against S1+S2+APD measurements. Bottom row: Results from the AF pilot study. (c), and (d): Two instances for a pilot study on predicting AF from parameters calibrated against the S1S2 protocol. (c) AF is terminated for the ground truth as well as all of the posterior samples; (d) AF is terminated for the ground truth, and all but 11 of the posterior samples.}\label{fig:mcmc:main}
    \end{minipage}
    
\end{figure}

We applied the proposed approach to a series of synthetic problems of calibration of tissue parameters of the left atrium and summarised the performance in Figures \ref{fig:par:scatter} and \ref{fig:err:boxplots}. A specific case is shown in Figure \ref{fig:lat:maps} for clarity. For a couple of instances, we show a comparison of the posterior distributions obtained via Algorithm \ref{alg:v3} with those obtained from MCMC, treating it as a gold standard in Figure \ref{fig:mcmc:main}; the ground truth parameter is also shown for reference. We then performed a pilot study to determine whether AF is sustained or not given measurements only from the S1S2 pacing protocol. As demonstrated with the pilot study, the proposed method shows promise to enable this prediction with reasonable accuracy (about 94\% posterior samples predicted the AF behaviour correctly) in real-time as we move the expensive training of GPEs offline.

Although demonstrated primarily in the context of cardiac electrophysiology, the proposed method may be readily adapted to a broad class of static inverse problems across diverse application domains. 
A natural extension is to calibrate spatially heterogeneous tissue parameters for more realistic cardiac modeling. Such fields can be represented using methods like Gaussian process manifold interpolation\cite{coveney2020gaussian} or basis expansions informed by imaging, though the optimal choice remains an open question. For example, high-resolution activation time measurements can be used to infer spatially varying coefficients, enabling the estimation of local conduction heterogeneities\cite{van2021identification}. Moreover, region-specific conduction slowing—such as in Bachmann’s bundle observed even in sinus rhythm\cite{heida2021reduction}—can be incorporated as spatially varying conductivities to guide parameter estimation.
Once parameterised, the coefficients can be inferred using Algorithm \ref{alg:v3}, with only a moderate increase in dimensionality. This is manageable with richer measurement data—clinical settings often provide hundreds of data points, making the extension both feasible and promising.
 
The current pseudo-dynamics, modeled as Brownian motion, do not inherently enforce physiological constraints on parameters (e.g., positivity or bounded ranges). While the EnKF update often corrects violations, proposals outside the trained GPE region—such as negative values—can lead to high emulator uncertainty and weak updates, slowing convergence. A promising direction is to incorporate constrained dynamics (e.g., reflecting or truncated Brownian motion) to ensure physically plausible proposals and improve numerical stability. Approaches include Doob’s h-transform\cite{roy2017stochastic}, killed diffusions \cite{kappen2005path}, or manifold-based Brownian motion with tailored metrics\cite{hsu2002stochastic,mamajiwala2022stochastic}.

Another key direction for future work is the design of clinical experiments\cite{rainforth2024modern} to improve calibration quality. This includes selecting informative measurements, influenced by stimulation protocols, sensor placement, and tissue properties. In our context, the goal is to identify a pacing strategy that best informs AF-related parameters. Optimising sensor number and spatial distribution—especially in heterogeneous tissue—can enhance parameter identifiability, with methods like sensitivity analysis or information-theoretic criteria offering guidance. Additionally, anatomical factors such as atrial shape, fiber orientation, and wall thickness affect measurements; incorporating insights from virtual cohorts or statistical shape models \cite{piazzese2017statistical} can improve robustness and generalisability.

\clearpage
\section*{Acknowledgements}
This work used the ARCHER2 UK National Supercomputing Service (https://www.archer2.ac.uk).

\bibliographystyle{plainnat}

\clearpage
\renewcommand{\thesection}{S\arabic{section}}
\setcounter{section}{0}
\setcounter{figure}{0}

{\Large \textbf{Supplementary material}}

 \section{Analysis: influence of $\sigma_{\theta}$}\label{sec:proof}

\renewcommand{\thefigure}{S\arabic{figure}}

\renewcommand{\thetable}{S\arabic{table}}

Although a detailed analysis of the proposed algorithm is beyond the scope of this work, we provide a sketch analysis of the effect of the additional parameter $\sigma_{\theta}$ on the posterior mean and variance obtained via Algorithm \ref{alg:v3}. Previous work has shown that when $h$ is linear, $\sigma_{\theta}=0$ and no emulator is used, Algortihm \ref{alg:v3} reproduces the true posterior mean and covariance\cite{schillings2017analysis}.

In the presence of an emulator, we have shown the convergence to the true posterior mean and variance with reducing emulator uncertainty (and ultimately in the absence of an emulator) with the help of a toy problem in Section \ref{sec:toy}. The main challenge, then, is to understand the influence of a non-zero $\sigma_{\theta}$ on the posterior mean and variance.  To explore this,  we draw upon the Kushner-Stratonovich equation\cite{roy2017stochastic}, alongside previous work\cite{schillings2017analysis} which analyzed Algorithm \ref{alg:v3} when $h$ is linear, $\sigma_{\theta}=0$, for a finite ensemble size and time-step size tending to 0.

Consider the Kushner-Stratonovich equation, which characterizes the evolution of the posterior expectation of functions of the system’s state and parameters, updating these expectations as new, noisy measurements become available.
 
By selecting appropriate functions, such as the identity function or quadratic functions, we can use the Kushner-Stratonovich equation (Equation (\ref{eqn:ks})) to derive the evolution equations for the first and second moments (hence mean and covariance) of the posterior distribution.
    \begin{equation}\label{eqn:ks}
        \pi_t (\phi)=   \pi_0 (\phi) + \int_0^t \pi_s(L_s(\phi))ds + \int_0^t (\pi_s(\phi h) -\pi_s(\phi )\pi_s(h) ) dI_s
    \end{equation}
Here, $\pi_t(\phi)$ represents the posterior expectation of $\phi(\theta_t)$ for any  $\phi \in L^2(\pi_t)$, and $L_s(\phi)$ is the generator of the process dynamics \cite{roy2017stochastic}. $I_t$ is the innovation, or new information, due to measurements $Y_t$, defined by $I_t=Y_t-\int_0^t \pi_s(h(\theta_s)) ds$. The first term in Equation (\ref{eqn:ks}) represents the effect of the initial distribution $\pi_0$, the second term accounts for the process dynamics, and the third term the measurements. To explore the impact of $\sigma_{\theta}$ on the posterior distribution, we focus on the second term, which relates to the process dynamics.

For the Brownian motion-based process dynamics, given by $d\theta_t=\sigma_{\theta}dB_t$, the generators for the first and second moments of $\theta$ are $L_s(\theta_s) = 0$ for $\phi(\theta_s)=\theta_s$, and  $L_s(\theta_s \theta_s^T) = \sigma_{\theta}\sigma_{\theta}^T$ for $\phi(\theta_s)=\theta_s \theta_s^T$. Consequently, the second term in Equation (\ref{eqn:ks}) simplifies to $\sigma_{\theta}\sigma_{\theta}^Tt$. This indicates that the artificial parameter dynamics lead to an overestimation of the posterior variance at time $t$ by $\sigma_{\theta}\sigma_{\theta}^Tt$.

This result is consistent with the evolution of posterior variance in the Kalman-Bucy filter\cite{bishop2023mathematical}, which provides the continuous-time solution for both posterior mean and variance under linear process and measurement dynamics. Since our process dynamics are linear, the effect of the process noise remains the same whether viewed from the perspective of the Kalman-Bucy filter (where the measurement function is also linear) or the Kushner-Stratonovich equation (where the measurement function is non-linear). Indeed, Section 6.3.3 of \cite{roy2017stochastic} demonstrates how the Kalman-Bucy equations are derived from the Kushner-Stratonovich equations when linear process and measurement dynamics are assumed.

Thus, we have shown that the posterior mean remains the same (since $L_s(\theta_s) = 0$) while the posterior variance obtained using Algorithm \ref{alg:v3} is inflated by $\sigma_{\theta}\sigma_{\theta}^Tt$. However, based on our extensive numerical experiments, we observed that this effect was negligible when compared to the posterior variance obtained from MCMC methods. It is important to note that $\sigma_{\theta}$ is often chosen to be a small value to facilitate particle perturbation, with its square being even smaller. Additionally, discretization—both in terms of the ensemble and over time—plays a significant role in the numerical accuracy.

\section{Sampling parameters to train a GPE}\label{sec:sampling}
To efficiently train GPEs for tissue-level outputs, a multi-stage parameter sampling and screening pipeline was used. The aim was to avoid unfeasible simulations and reduce compuatational cost while ensuring coverage of the parameter space. The key steps are outlined below: 
\begin{itemize}
\item \textbf{Cell-level simulations:} Simulate single-cell models using only the four intrinsic ionic parameters: $\tau_{in}, \tau_{out}, \tau_{open}, \tau_{close}$
\item \textbf{Feasibility labelling:} Classify each sampled set of parameters as feasible or unfeasible based on simulation outputs. A parameter set is labelled as unfeasible if it results in alternans or if the action potential duration (APD) exceeds 500\,ms.
\item \textbf{Logistic regression classifier:} Fit a logistic regression classifier using the four parameters as inputs and labels from step 2 as targets.

\item \textbf{Tissue-level parameter sampling:} Generate a Latin Hypercube Sample (LHS) of 5-dimensional parameter vectors covering the full tissue parameter space (including conductivity) within the bounds listed in Table~\ref{tab:tissue} \cite{corrado2018work}.

\item \textbf{First rejection stage:} First rejection of the LHS is based on logistic regression classifier (from step 3). Since the classifier is based on 4 parameters, it is applied on the corresponding 4-dimensional subset of the LHS. 

\item \textbf{Left atrium simulations:} For the remaining parameter sets, simulate an S1S2 protocol on the entire left atrium. The electrophisiology was simulated with openCARP\cite{openCARP-sw} on the ARCHER2 supercomputing facility.

\item \textbf{Second rejection stage:} Discard any parameter sets that still result in unfeasible tissue outputs despite passing the first rejection stage.
\end{itemize}
Starting with an initial LHS of 350 parameter sets, 202 remained after both rejection stages. GPEs were trained for 
176 of the 202 feasible samples and the remaining 26 were left for validation. A separate GPE was trained for each output type and each measurement location resulting in 45 GPEs. The inputs used for training are the 5 tissue parameters whereas the outputs are as described in Section \ref{sec:calib:la}. To train every GPE, the prior mean function was chosen to be linear, and the prior covariance was modelled using a radial basis function (RBF) kernel. Hyperparameters of the mean and kernel functions were then estimated by maximizing the marginal log-likelihood.

\begin{table}[htbp]
\caption{\label{tab:tissue} Ranges of mMS paramters for Latin hypercube sampling}
\vspace{4 mm}
\centerline{\begin{tabular}{lccccc} \hline\hline
 & $\tau_{in}$ & $\tau_{out}$ &$\tau_{open}$ & $\tau_{close}$ & conductivity \\ 
 &&&&& ($cm^2/s$) \\
$min$ & 0.01 &1 &65&100 &0.1\\
 $max$& 0.3& 30& 215 & 150 & 5\\ \hline\hline
\end{tabular}}
\end{table}

\section{Non-linear toy problem}\label{sec:toy}
\begin{figure}[t]
    \centering
    \includegraphics[width=\linewidth]{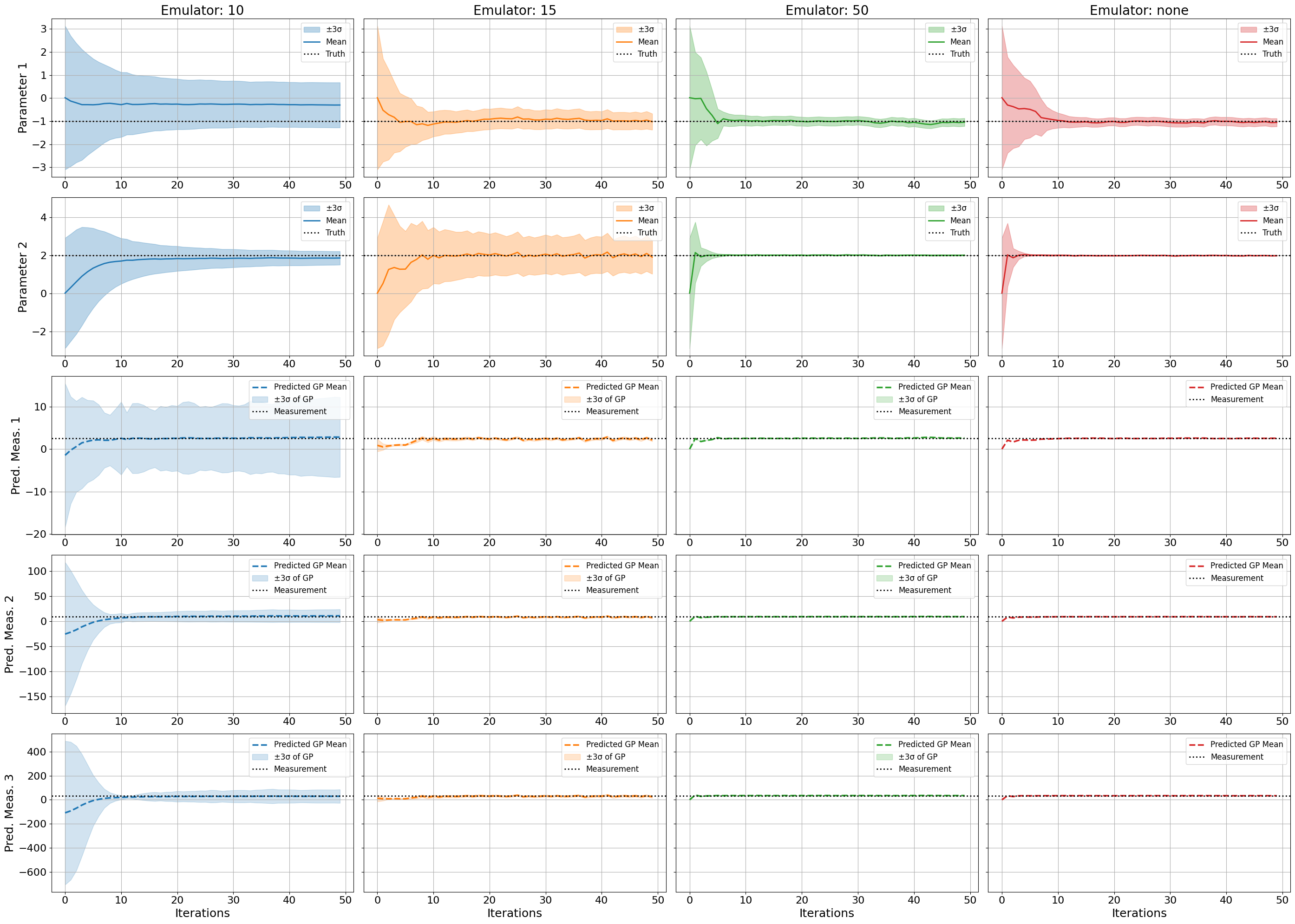}
    \caption{Comparison of EnKF with and without emulator. First 3 columns correspond to emulators with different number of points used to train them, i.e. 10, 15 and 50. Last column corresponds to results with the known measurement function. First two rows correspond to the evolution of parameters with iterations along with the 3$\sigma$ bounds of the ensemble. Last three rows compare the true measurement with that predicted from the ensemble mean at every iteration - the 3$\sigma$ bounds here correspond the GPE standard deviation.}
    \label{fig:toy}
\end{figure}

We first consider the following toy problem:
\begin{equation}\label{eqn:toy}
    y = -\theta_1^3 x+ \theta_2^3 x^2
\end{equation}
Here $x$ is similar to a location variable. We assume a ground truth value for $[\theta_1,\theta_2] \equiv [-1.5,2.0]$ and create a noisy measurement vector for 3 locations $x \equiv [0.5,1.0,2.0]$ by adding a zero-mean Gaussian noise with standard deviation $\sigma_y=0.05\mathbb{I}_{3 \times 3}$. We then train several GPEs (using the GaussianProcessRegressor library in sklearn\cite{hao2019machine} in python) for $y$ as a function of the two parameters in the range $[-5,5]$ by choosing 10, 15 and 50 training points for each of the measurement locations. Starting with an initial ensemble (of size N=500) drawn from a 2-dimensional standard Gaussian distribution, we use Algorithm \ref{alg:v3} for 50 (pseudo-)time steps with $\Delta t=0.01sec$ to solve the calibration problem. Results for this toy problem are presented in Figure \ref{fig:toy}. Consider the first column, the GPE predictive mean of at the ensemble mean is close to the true measurement, even though the parameters are away from the truth - this is due to the poor quality of the emulator. The Algorithm is agnostic to the true values of the function and only relies on the GPE mean predictions. Therefore, moving from left to right we can see the parameter accuracy improving with the improved emulators even though the predicted measurements at the ensemble means consistently match the true measurements. We also observe, going from left to right that the posterior variance goes on reducing, and the posterior variance corresponding to the emulator trained with 50 points and those without an emulator (i.e. the last but one and last columns) are almost identical.

\section{Scaling the algorithmic measurement noise intensity}\label{sec:scalesig}
 Let the measurement noise intensity be $\sigma_y$, the likelihood may then be written as 
 $$\pi(y|\theta) = \exp\left(-\frac{(y-h(\theta))^2}{2\sigma_y^2}  \right)$$
Let us now suppose the same measurement is available $N$ times such that $y_1 = y_2 = ... = y_K = y$ instead, the likelihood then becomes
 $$ \pi(y_1,y_2...y_K|\theta) = \prod_{i=1}^K  \pi(y_i|\theta) = \prod_{i=1}^K  \exp\left(-\frac{(y_i-h(\theta))^2}{2\sigma_y^2}  \right) = \exp\left(-K \frac{(y-h(\theta))^2}{2\sigma_y^2}  \right)$$
Thus resulting in a variance that is $\sigma_y^2/N$ (assuming $y_i=y \; \forall i$). Hence, to ensure that 
$K$ perturbed measurements collectively represent the same information as a single measurement with variance $\sigma_y^2$
 we scale the variance of each individual perturbation by $K$. 
 
\begin{figure}[b]
    \centering
    \begin{subfigure}{0.25\linewidth}
        \includegraphics[height=1.2in]{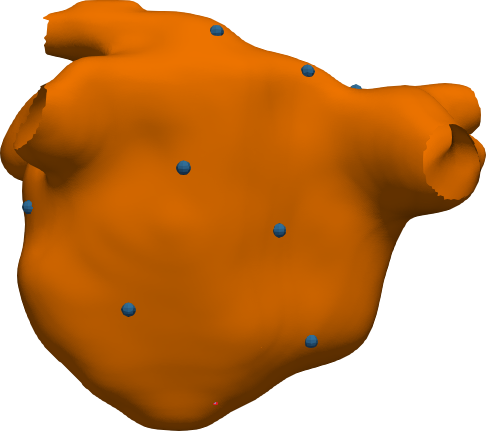}
    \caption{}
    \end{subfigure}
        \begin{subfigure}{0.25\linewidth}
        \includegraphics[height=1.2in]{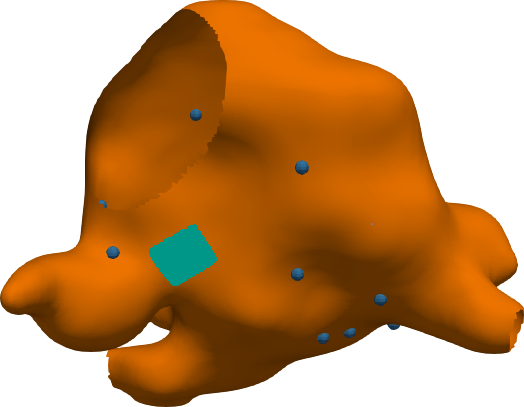}
    \caption{}
    \end{subfigure}
    \caption{Two views of the left atrium anatomy illustrating the 15 measurement locations (blue dots) and stimulus region (green patch) for the S1S2 pacing protocol.}
    \label{fig:stim}
\end{figure}

\end{document}